\documentclass[useAMS,usenatbib]{mnras}
\usepackage{graphicx}
\usepackage[outdir=./]{epstopdf}
\usepackage{tablefootnote}
\usepackage{gensymb}
\DeclareGraphicsExtensions{.eps}

\renewcommand{\d}{\mathrm{d}}

\newcommand{\Msun}{\mbox{$\mathrm{M}_{\odot}$}}
\newcommand{\Rsun}{\mbox{$\mathrm{R}_{\odot}$}}

\title[Variability of the gaseous disc at SDSS\,J1043+0855]{Another one grinds the dust: Variability of the planetary debris disc at the white dwarf SDSS\,J104341.53+085558.2}

\author[Manser et al.]{Christopher J. Manser\,$^1$\,{\Huge \footnotemark},
Boris T. G\"ansicke\,$^1$,
Detlev Koester\,$^2$,
\newauthor Thomas R. Marsh\,$^1$,
John Southworth\,$^3$\\
$^{1}$ Department of Physics, University of Warwick, Coventry CV4 7AL,
UK \\
$^{2}$ Institut f\"ur Theoretische Physik und Astrophysik, University of Kiel,
24098 Kiel, Germany\\
$^{3}$ Astrophysics Group, Keele University, Staffordshire, ST5 5BG, UK
}

\begin{document}

\date{Accepted 20XX. Received 20XX; in original form 20XX}

\pagerange{\pageref{firstpage}--\pageref{lastpage}} \pubyear{20XX}

\maketitle

\label{firstpage}

\begin{abstract}

We report nine years of optical spectroscopy of the metal-polluted white dwarf SDSS\,J104341.53+085558.2, which presents morphological variations of the line profiles of the 8600\,\AA\ Ca\,{\textsc{ii}} triplet emission from the gaseous component of its debris disc. Similar changes in the shape of the Ca\,{\textsc{ii}} triplet have also been observed in two other systems that host a gaseous disc, and are likely related to the same mechanism. We report the Mg, Si, and Ca abundances of the debris detected in the photosphere of SDSS\,J1043+0855, place upper limits on O and Fe, and derive an accretion rate of $(2.5$ - $12)\,\times\,10^{8}$\,gs$^{-1}$, consistent with those found in other systems with detected debris discs. The Mg/Si ratio and the upper limit on the Fe/Si ratio of the accreted material broadly agree with those found for the crust of the Earth. We also review the range of variability observed among white dwarfs with planetary debris discs.

\end{abstract}


\footnotetext{E-mail: C.Manser@Warwick.ac.uk}

\section{Introduction}

The detection of metal pollution in white dwarf atmospheres provides strong evidence that 25-50\,per\,cent of white dwarfs host remnants of planetary systems \citep{zuckermanetal03-1, koesteretal14-1, koester+kepler15-1}. The survival of planets through the post main-sequence evolution of their host star is further supported by the detection of more than 35 dusty debris discs at white dwarfs with metal pollution \citep{kilicetal06-1, farihietal08-1, farihietal09-1, juraetal09-1, debesetal11-2, hoardetal13-1, bergforsetal14-1, rocchettoetal15-1}, and is also predicted by theoretical studies \citep{villaver+livio07-1, veras+gaensicke15-1}.

Debris discs around white dwarfs are thought to be produced by the tidal disruption of asteroids (or comets) scattered onto a highly eccentric orbit by planets in the system \citep{grahametal90-1, jura03-1}. While recent simulations have begun to explore the formation and evolution of these discs; from the initial disruption of the planetesimal and the shrinking of its orbit \citep{debesetal12-1,verasetal14-1, verasetal15-1}, to the dynamics and interactions of the debris within the disc \citep{rafikov11-2, metzgeretal12-1}, many aspects remain poorly understood. 

Over the last decade, gaseous debris discs have been detected around eight white dwarfs, all of which also host circumstellar dust and have metal polluted photospheres \citep{gaensickeetal06-3, gaensickeetal07-1, gaensickeetal08-1, gaensicke11-1, farihietal12-1, melisetal12-1, wilsonetal14-1, guoetal15-1}. These systems were identified by the detection of Ca\,{\textsc{ii}} 8498.02\,\AA, 8542.09\,\AA, 8662.14\,\AA\ emission lines (henceforth the Ca\,{\textsc{ii}} triplet, air wavelengths are given).

\begin{figure*}
\centerline{\includegraphics[width=2\columnwidth]{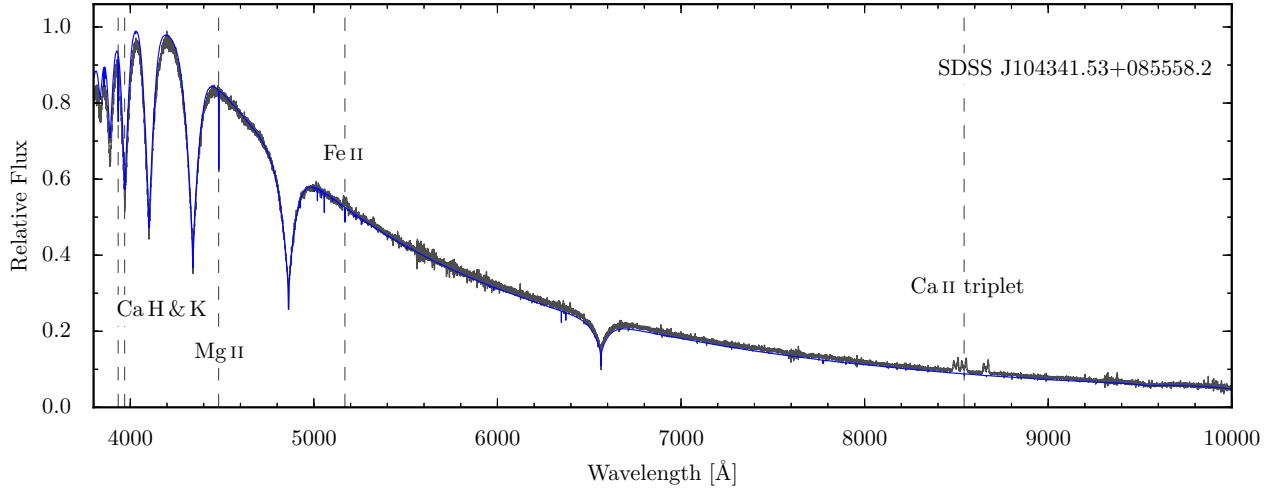}}
\caption{\label{f-1043_plot} X-Shooter spectra of SDSS\,J1043+0855 (grey, obtained in January and May 2011) together with a model atmosphere fit (blue) using the atmospheric parameters listed in Table\,\ref{t-wds}. The strongest emission and absorption lines have been labelled (note that the Fe\,{\textsc{ii}} feature is in emission).}
\end{figure*} 

\begin{table*}
\centering
\caption{Log of observations of SDSS\,J1043+0855. $^{a}$ Different exposure times for the individual X-Shooter arms (UVB / VIS). We did not use data collected by the NIR arm of X-Shooter as the signal-to-noise ratio was too poor. \label{t-dates}}

\begin{tabular}{rrrrr}
\hline
Date & Telescope/Instrument  & Wavelength Range [\AA] & Resolution [\AA] & Exposure Time [s]\,$^{a}$ \\
\hline
2003 April 05 & SDSS & 3800 -- 9200 & 2.9 & 2900\\
2007 February  03 & WHT/ISIS & 7400 -- 9200 & 2.0 & 4800\\
2009 February 16 & WHT/ISIS & 6000 -- 8900 & 3.7 & 3950\\
2010 April 22 & WHT/ISIS & 8100 -- 8850 & 1.1 & 7200\\
2011 January 29 & VLT/X-Shooter  & 2990 -- 10400 & 1.12 & 2950 / 2840\\
2011 May 30 & VLT/X-Shooter  & 2990 -- 10400 & 1.15 & 2950 / 2840\\
2012 January 03 & SDSS & 3602 -- 10353 & 3.2 & 2702\\

\hline
\end{tabular}
\end{table*}

So far only four of these eight gas disc systems have multi-epoch spectroscopy over time scales of years, three of which show variability in the Ca\,{\textsc{ii}} emission of the gaseous disc; either as significant changes in the morphology of the Ca\,{\textsc{ii}} triplet \citep{wilsonetal15-1, manseretal16-1}, or as a decrease in strength of the lines \citep{wilsonetal14-1}. It is thought that the gaseous components of the debris discs in these systems are tracers of dynamic activity \citep{gaensickeetal08-1, wilsonetal14-1, wilsonetal15-1, manseretal16-1}.

The gaseous disc around SDSS\,J104341.53+085558.2 (henceforth SDSS\,J1043+0855) was discovered by \cite{gaensickeetal07-1} via the detection of Ca\,{\textsc{ii}} triplet emission. An infrared excess was detected by \cite{melisetal10-1} and \cite{brinkworthetal12-1}, confirming the presence of a dusty disc in SDSS\,J1043+0855. Here, we report nine years of spectroscopic observations of SDSS\,J1043+0855 that reveal a change in the morphology of the Ca\,{\textsc{ii}} triplet, similar to those seen in other gaseous disc systems (SDSS\,J122859.93+104032.9 and SDSS\,J084539.17+225728.0, henceforth SDSS\,J1228+1040 and SDSS\,J0845+2257, \citealt{wilsonetal15-1, manseretal16-1}). We also present the accretion rates of the debris onto the white dwarf and the metal abundances of the debris.

\section{Observations}
\label{sec:observations}

We obtained optical spectroscopy of SDSS\,J1043+0855 from 2003 to 2012 with several instruments: X-Shooter \citep{vernetetal11-1} on the ESO Very Large Telescope (VLT); the 2.5\,m Sloan Digital Sky Survey telescope (SDSS, data retrieved from DR7 and DR9, \citealt{gunnetal06-1, abazajianetal09-1, eisensteinetal11-1, ahnetal14-1, smeeetal13-1}); and the Intermediate dispersion Spectrograph and Imaging System (ISIS) on the William Herschel Telescope (WHT). A log of the observations is summarised in Table\,\ref{t-dates}. The X-Shooter data were reduced within the \texttt{\textsc{reflex}}\,\footnote{Documentation and software for \texttt{\textsc{reflex}} can be obtained from http://www.eso.org/sci/software/reflex/} reduction work flow using the standard settings and optimising the slit integration limits \citep{freudlingetal13-1}. The sky spectrum of each observation was used to determine the spectral resolution. The first ISIS spectrum of SDSS\,J1043+0855 was reported in \cite{ gaensickeetal07-1}, the additional ISIS spectra were obtained with a similar setup, and were reduced in the same fashion (see \citealt{farihietal12-1} and \citealt{wilsonetal14-1} for additional details). We removed the telluric lines present in the VIS arm of the X-Shooter spectra using the X-Shooter Spectral Library (XSL) provided by \cite{chenetal14-1}, and applying the method outlined in \cite{manseretal16-1}.

\begin{figure}
\centerline{\includegraphics[width=0.95\columnwidth]{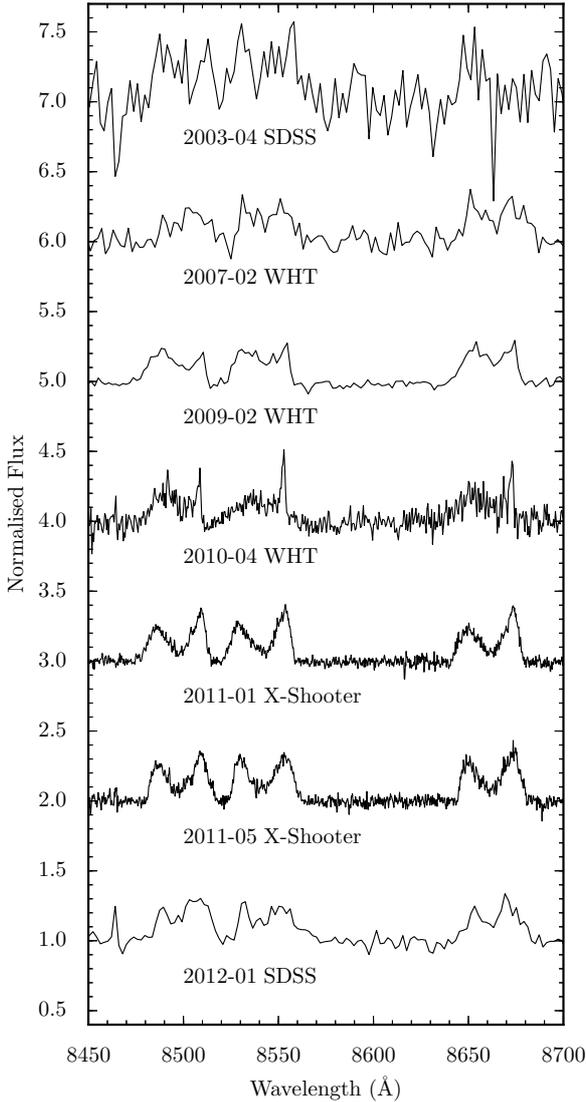}}
\caption{\label{f-1043_Ca_vel} Time series spectroscopy of the Ca\,{\textsc{ii}} triplet of SDSS\,J1043+0855 spanning nine years. The spectra are normalised and offset in steps of one from the 2012 January observation.}
\end{figure}

\section{Evolution of the calcium emission profile}

\subsection{Double peaked emission from a disc}

The characteristic emission profile from a gaseous disc with a radially symmetric intensity distribution and circular orbit is a symmetric, double-peaked emission profile (see Figure 1 of \citealt{horne+marsh86-1}). The width of the profile arises from the wide range of velocities projected along the line of sight and can reveal details about the structure and geometry of the gaseous disc. The largest velocities correspond to material at the inner disc radius, which can hence be determined from the full width at zero intensity, i.e. the point at which the emission drops to the continuum level, and with knowledge of the inclination, $i$, using $R\,\sin^{2}\,i = R_{\textrm{obs}}$, where $R$ and $R_{\textrm{obs}}$ denote the actual and observed radii respectively. We define the maximum red/blue-shifted velocities to represent the red/blue inner edges of the disc. The outer radius can be estimated from the peak separation in the double peaked profile.

Any departure from a radially symmetric, circular disc will manifest itself as asymmetries in the double peaked emission profile. An eccentric disc of uniform intensity, for example, would generate an asymmetric double-peaked profile if viewed along the semi-minor axis, as there is an asymmetry in the velocity distribution of the material, as well as in the total red and blue-shifted light emitted from the disc. We use this insight below to discuss the changes in the line profiles observed at SDSS\,J1043+0855. 

\subsection{Variation of the Calcium emission lines}

The Ca\,{\textsc{ii}} triplet in SDSS\,J1043+0855 changes in morphology over a time scale of nine years (Figure\,\ref{f-1043_Ca_vel}). All three components of the triplet vary in the same manner and are henceforth referred to in singular. 

The initial three spectra are noisy and of low resolution, showing a broad line profile. In contrast, finer features appear in the higher resolution 2010 WHT spectrum, namely a sharp red-shifted peak and a gradual drop off to blueward wavelengths, revealing a clear asymmetry in the red and blue inner edges of the disc. This asymmetry decreases in the 2011 January spectrum, and vanishes in the 2011 May spectrum. The second SDSS spectrum obtained in 2012 shows no sign of any sharp departures from symmetry, although it is of relatively low spectral resolution. In Section\,\ref{sec:discussion} we discuss the similarities, and the possible origin, of the observed variations with those seen in other systems.

\subsection{Inclination of the disc}

In the 2011 January and May spectra, the `valley' in between the two peaks of the emission profiles almost reaches the continuum level, which suggests that the debris disc is optically thick and seen at a large inclination. \cite{horne+marsh86-1} have shown that in an optically thick accretion disc, line emission is more likely to escape along paths of largest velocity gradient provided by Keplerian shear flow, which is at a minimum for purely tangential or radial emission through the disc. For an observer looking through the disc at an inclination of 90$^{\circ}$ (edge on), material travelling perpendicular to the line of sight (emission in the `valley') will be emitting along the radial direction of the disc and will therefore be suppressed.

\begin{figure}
\centerline{\includegraphics[width=1\columnwidth]{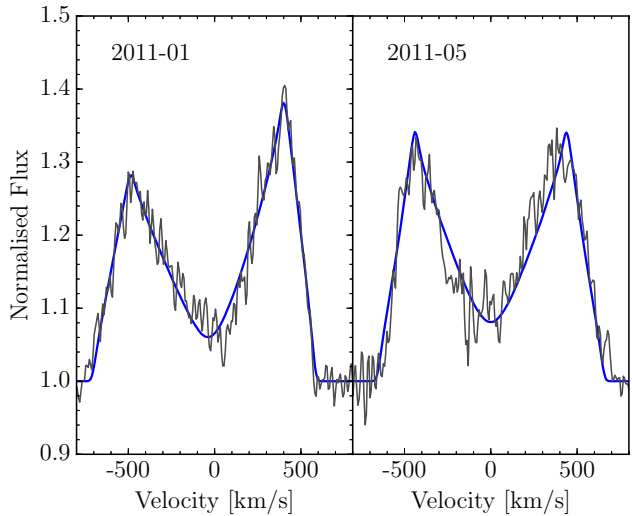}}
\caption{\label{f-fit} Model fit to the normalised 8542.09\,\AA\ component of the Ca\,{\textsc{ii}} triplet for the 2011 January and May X-shooter spectra, from which we obtain an inclination of $i \simeq 74^{\circ}$ for the disc at SDSS\,J1043+0855. The model applies the formulation described in \protect\cite{horne+marsh86-1} which has been expanded to include eccentric orbits (see Appendix\,\ref{appendix-1} for more details).}
\end{figure}

While the formulation described in \cite{horne+marsh86-1} was developed for circular orbits, it was expanded to include eccentric orbits in order to model the emission profile of SDSS\,J1228+1040 obtained in 2006 \citep{gaensickeetal06-3}, and we give additional details on this extension in Appendix\,\ref{appendix-1}. We fit the emission profiles of the 2011 January and May spectra of SDSS\,J1043+0855 in the same manner (see Figure\,\ref{f-fit}), and obtain an inclination of $i\simeq$\,76$^{\circ}$ and 72$^{\circ}$ respectively, leading to an average inclination of $i=$\,74$^{\circ}$. The statistical uncertainty obtained for the two inclinations was $\pm$\,0.5$^{\circ}$, which is probably an underestimate, and a more realistic uncertainty is taken to be $\simeq$\,$\pm$\,5$^{\circ}$.

\section{Metal abundances in the photosphere of SDSS\,J1043+0855}
\label{sec:abund}

We fitted the two SDSS spectra with DA (hydrogen dominated) white dwarf model atmospheres using the methods described in \cite{gaensickeetal12-1} and \cite{koesteretal14-1}, and find $T_{\textrm{eff}}$ = 17879\,$\pm$\,195\,K, and $\log$\,$g$ = 8.124\,$\pm$\,0.033, corresponding to $M_{\textsc{wd}}$ = 0.693\,$\pm$\,0.020\,\Msun\ and $R_{\textsc{wd}}$ = 0.0120\,$\pm$\,0.0003\,\Rsun\ (see Table\,\ref{t-wds}). Using the initial-to-final mass relation of \cite{casewelletal09-1}, \cite{kaliraietal08-1}, \cite{catalanetal08-1}, and \cite{williamsetal09-1}, we estimate the mass of the white dwarf progenitor to be 2.76\,$\pm$\,0.04\,\Msun.

We note that while the atmospheric parameters derived here are consistent with previous measurements \citep{eisensteinetal06-1, gaensickeetal07-1, tremblayetal11-2, kleinmanetal13-1} the $ugriz$ photometry suggests a lower effective temperature, even when allowing for the maximum reddening along the line of sight obtained from \cite{schlafly+finkbeiner11-1}. This discrepancy is likely due to a higher amount of extinction, either by a denser patch in the interstellar medium that remains unresolved in the dust maps, or by circumstellar dust. We speculate that there could be additional dust in the system, which may not be associated with the dusty component of the debris disc at SDSS\,J1043+0855 \citep{melisetal10-1, brinkworthetal12-1}, and could possibly be the cause of the observed reddening.

\begin{table*}
\centering
\caption{Metal polluted white dwarfs with circumstellar gas detected in emission (e) or absorption (a), and evidence for photometric or spectroscopic variability (v). System parameters and accretion rates are given with errors where known. Values derived or updated in this paper are set in italics.
$^{1}$ \protect\cite{dufouretal12-1}, 
$^{2}$ \protect\cite{wilsonetal15-1}, 
$^{3}$ \protect\cite{farihietal12-1}, 
$^{4}$ \protect\cite{xu+jura14-1}, 
$^{5}$ \protect\cite{gaensickeetal07-1}, 
$^{6}$ \protect\cite{melisetal10-1}, 
$^{7}$ This paper, 
$^{8}$ \protect\cite{guoetal15-1}, 
$^{9}$ \protect\cite{gaensickeetal06-3}, 
$^{10}$ \protect\cite{gaensickeetal12-1}, 
$^{11}$ \protect\cite{koesteretal14-1}, 
$^{12}$ \protect\cite{manseretal16-1}, 
$^{13}$ \protect\cite{koesteretal05-2}, 
$^{14}$ \protect\cite{vossetal07-1}, 
$^{15}$ \protect\cite{melisetal12-1}, 
$^{16}$ \protect\cite{wilsonetal14-1}, 
$^{17}$ \protect\cite{vennes+kawka13-1},
$^{18}$ \protect\cite{koester+wilken06-1}, 
$^{19}$ \protect\cite{debesetal12-1}, 
$^{20}$ \protect\cite{vanderburgetal15-1}, 
$^{21}$ \protect\cite{xuetal16-1}. \label{t-wds}}
\begin{tabular}{lrllllllr}
\hline
Name & Type & $\log\,g$ & $T_{\textrm{eff}}$ & $M_{\mathrm{WD}}$ & $\tau_{\textrm{cool}}$ & $\dot{M}$ & Features & ref \\
 &  & (g\,cm$^{-2}$) & (K) &  (\Msun) & (Myr) & ($\times\,10^{8}$\,g\,s$^{-1}$) & &  \\
\hline
SDSS\,J0738+1835  & DB & 8.4 (0.2)          & 13950 (100) & 0.841 (0.131)    & 477 (160) & $1300$             & e  & 1 \\
SDSS\,J0845+2257  & DB & 8.18 (0.20)     & 19780 (250) & \textit{0.73 (0.11)}  & \textit{122 (44)}    & $160$           & e, v  & 2 \\
SDSS\,J0959--0200  & DA & 8.06 (0.03)     & 13280 (20)   & 0.64 (0.02)        & \textit{324 (17)}               & $0.32$               & e, v  & 3, 4 \\
SDSS\,J1043+0855  & DA & \textit{8.124 (0.033)} & \textit{17879 (195)} & \textit{0.693 (0.020)}  & \textit{153 (10)}          & \textit{(2.5 - 12)}      & e, v  & 5, 6, 7 \\
WD\,1144+0529      & DA & 7.74 (0.03)     & 23027 (219) & 0.49 (0.03)       & 21.2 (1.9) & -                         & e  & 8 \\
SDSS\,J1228+1040  & DA & 8.150 (0.089) & 20713 (281) & 0.705 (0.051)  & 100 (5)     & $5.6$                  &  e, a, v & 6, 9, 10, 11, 12 \\
HE\,1349--2305       & DA & 8.133             & 18173          & 0.673               & 149.4       & $1.3$                   & e  & 13, 14, 15\\
SDSS\,J1617+1620  & DA & 8.11 (0.08)     & 13520 (200)  & 0.68 (0.05)      & 350 (50)   & $(6.4$ - $7.8)$    & e, v  & 16 \\
\hline
PG\,0843+516         & DA & 7.902 (0.089) & 22412 (304)  & 0.577 (0.047)  & \textit{42 (4)}               & $10.2$              & a  & 11 \\
WD\,1054--226           & DA & 8.04 (0.03)     & 7903 (16)      & -                       & \textit{1255 (92)}         & -                   & a & 17 \\
WD\,1124--293        & DA & 8.1                  & 9700            & 0.66                 & \textit{843}               & $1.3$                  & a  & 18, 19 \\
WD\,1145+017        & DB & -                     & 15900 (500)  & -                       & 175 (75)  & $430$                 & a, v  & 20, 21\\
\hline
\end{tabular}
\end{table*}

\begin{table*}
\centering
\caption{Diffusion timescales, $\tau_{\mathrm{diff}}$, and average accretion fluxes, $\dot{M}$, for the metals detected in the photosphere of SDSS\,J1043+0855. Photospheric abundances by number are given with respect to hydrogen. $^{a}$ We infer a total accretion rate based on the mass fluxes assuming a bulk Earth composition and the respective mass fractions for each element (from \citealt{allegreetal01-1}). \label{t-abund}}
\begin{tabular}{lrrrrr}
\hline
Element & $\log\,[\mathrm{Z}/\mathrm{H}]$ & $\tau_{\mathrm{diff}}$ [h] & $\dot{M}$ [g\,s$^{-1}$] & Bulk Earth mass fraction [per\,cent] $^{a}$ & Inferred total $\dot{M}$ [g\,s$^{-1}$] \\
\hline
8 O & $< -4.00$ & 65.0 & $< 2.0 \times 10^{9}$ & 32.4 & $< 6.2\times10^9$ \\
12 Mg & $-5.15$ (0.15) & 26.5 & $4.0 \times 10^{7}$ & 15.8 & $2.5\times10^8$\\
14 Si & $-4.80$ (0.15) & 13.8 & $2.0  \times 10^{8}$ & 17.1 & $1.2\times10^{9}$\\
20 Ca & $-6.00$ (0.20) & 18.6 & $1.3  \times 10^{7}$ & 1.6 & $8.0\times10^{8}$\\
26 Fe & $< -4.70$ & 11.1 & $< 6.3  \times 10^{8}$ & 28.8 & $< 2.2\times10^{9}$ \\
\hline
\end{tabular}
\end{table*}

\begin{figure}
\centerline{\includegraphics[width=1\columnwidth]{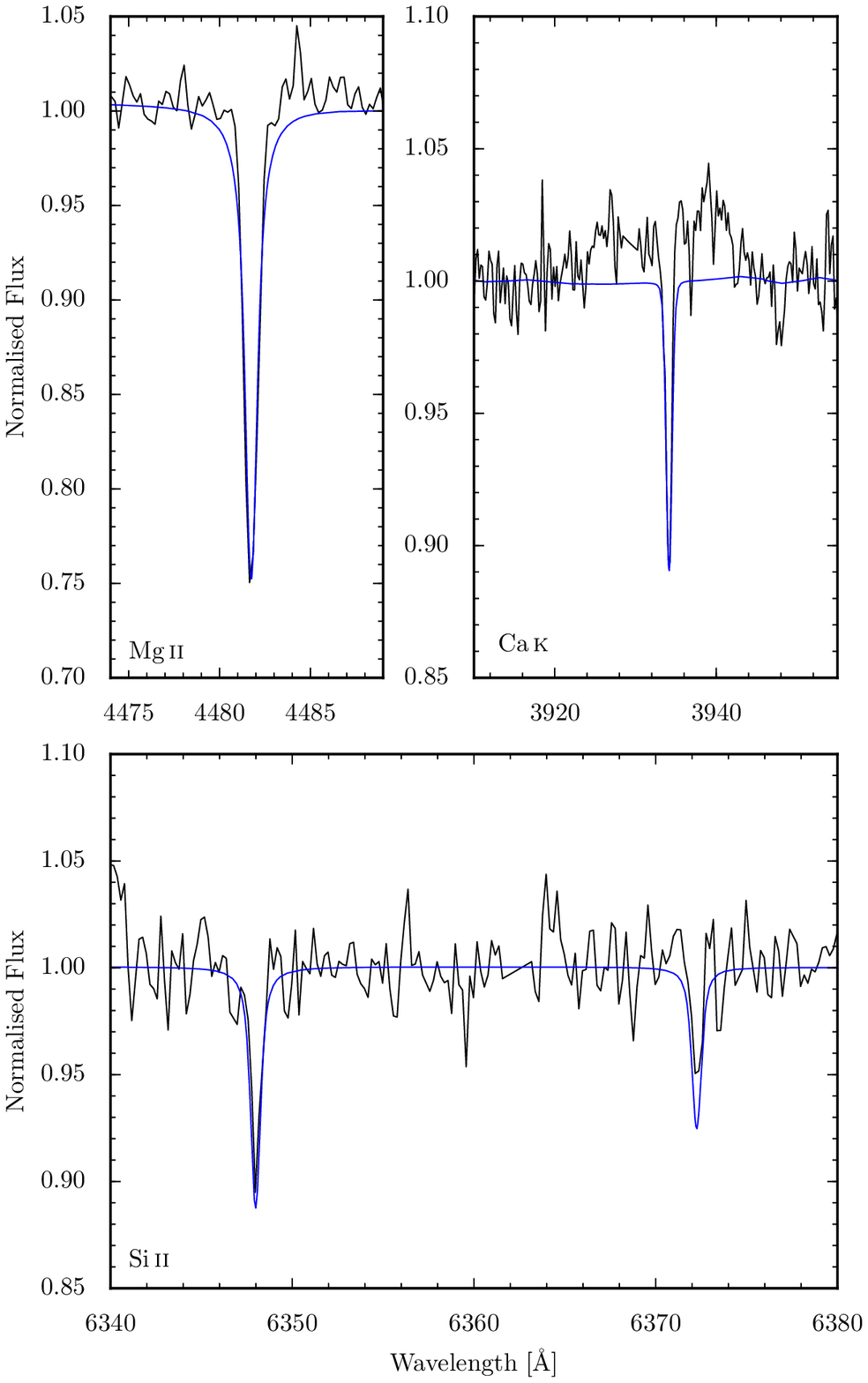}}
\caption{\label{f-absorp_lines} Model fits (blue) to absorption lines present in the combined X-Shooter spectra of SDSS\,J1043+0855 (black). The Ca\,{\textsc{k}} feature also contains a clearly detected emission profile from the gaseous disc.}
\end{figure}

Using the system parameters, the metal absorption lines detected in the spectra of SDSS\,J1043+0855 were modelled (see Figure\,\ref{f-absorp_lines}) to measure abundances relative to hydrogen which are given in Table\,\ref{t-abund}, along with the diffusion time scales and accretion fluxes. The abundance by number of Mg, Ca, and an upper limit for Fe relative to Si were found to be $\log$\,(Mg/Si) = $-0.64\,\pm\,0.21$, $\log$\,(Ca/Si) = $-1.33\,\pm\,0.24$, and $\log$\,(Fe/Si) $\leqslant 0.19$. From Figure 7 of \cite{jura+young14-1}, the ratios of Mg and Fe to Si are broadly consistent with those found for the crust of the Earth, and imply the accreted object is processed rather than having a 'chondritic' composition \citep{zuckermanetal07-1, xuetal13-1, xuetal14-1}. This is also supported by the relatively low number abundance of Mg with respect to Ca, $\log$\,(Mg/Ca) = $0.70\,\pm\,0.25$, when compared to a sample of 60 externally polluted white dwarfs (see Figure 1 of \citealt{jura+xu13-1}). Similarly low $\log$\,(Mg/Ca) ratios have also observed at other white dwarfs that are thought to accrete crust material, such as NLTT 43806 and NLTT\,19868 \citep{zuckermanetal11-1,kawka+vennes16-1}.

We also note that the Ca\,K 3934\,\AA\ feature shows clear emission in the averaged, normalised X-Shooter spectra (Figure\,\ref{f-absorp_lines}). Similar emission is seen at SDSS\,J1228+1040, which also revealed a difference in morphology between the Ca\,H\,\&\,K profiles and the Ca\,{\textsc{ii}} triplet, although the signal-to-noise of the current SDSS\,1043+0855 spectra is not high enough to determine this \citep{manseretal16-1}. 

We estimate the total accretion rate for each element from the mass flow rates given in Table\,\ref{t-abund}, assuming a bulk Earth composition and the respective mass fractions of each element (from \,\citealt{allegreetal01-1}). The resulting range from Mg, Si, and Ca, $\dot{M}_{\mathrm{Total}} = 2.5 \times 10^8 - 1.2 \times 10^9$\,gs$^{-1}$, reflects the uncertainty in the bulk abundances of the planetary debris, but agrees well with the distribution of accretion rates inferred for 19 other metal-polluted DA white dwarfs with dusty discs \citep{bergforsetal14-1}. For reference, the total accretion rate of SDSS\,J1228+1040 (where accretion fluxes for all major elements were measured from \textit{HST}/COS ultraviolet spectra) is $5.6\times10^{8}\,\textrm{g\,s}^{-1}$ \citep{gaensickeetal12-1}.

\begin{table}
\centering
\caption{Equivalent width measurements of the Ca\,{\textsc{ii}} triplet in SDSS\,J1043+0855. The errors given here are purely statistical. \label{t-ewtrip}}
\begin{tabular}{lr}
\hline
Date & Equivalent width [\AA] \\
\hline
2003--04 & -27 (3)\\
2007--02 & -22 (1)\\
2009--02 & -15 (1)\\
2010--04 & -13 (1)\\
2011--01 & -18 (1)\\
2011--05 & -20 (1)\\
2012--01 & -19 (1)\\
\hline
\end{tabular}
\end{table}

\section{Discussion}
\label{sec:discussion}

Up until the last decade, debris discs around white dwarfs have appeared static in nature, with no significant detections of variability in the properties of the disc itself or in the strength of the absorption lines in metal polluted systems. Due to the short diffusion time scales at the majority of these systems \citep{koesteretal14-1}, any change in the strength of the absorption lines would imply a change in the accretion rate onto the white dwarf. \cite{vonhippel+thompson07-1} claimed changes in the equivalent width of the photospheric Ca\,{\textsc{ii}}\,K line over a time scale of days at GD\,29--38, however additional observations obtained by \cite{debes+lopez-morales08-1} did not confirm such variation, and they concluded further data were required to determine the possible variable nature of the accretion onto the white dwarf.

In recent years, the spectroscopic monitoring of gaseous discs has revealed variability that gives us insight into their formation and dynamics. Table\,\ref{t-wds} lists the stellar parameters of the eight published gas disc systems. Other than SDSS\,J1043+0855, there are four gaseous disc systems with multi-epoch spectroscopy: SDSS\,J161717.04+162022.4 (\citealt{wilsonetal14-1}, henceforth SDSS\,J1617+1620), SDSS\,J0845+2257 \citep{gaensickeetal08-1, wilsonetal15-1}, SDSS\,J1228+1040 \citep{manseretal16-1}, and SDSS\,J073842.56+183509.6 (\citealt{gaensicke11-1, dufouretal12-1}, henceforth SDSS\,J0738+1835). 

SDSS\,J0845+2257, SDSS\,J1228+1040, and SDSS\,J1617+1620 all show variations on a time scale of years, but follow two distinct  types of evolution. In SDSS\,J1617+1620 the  Ca\,{\textsc{ii}} triplet emission gradually decreased in strength over a time scale of eight years while not undergoing noticeable changes in the line profile shape \citep{wilsonetal14-1}. In contrast, the changes seen in SDSS\,J0845+2257 and SDSS\,J1228+1040 are of a morphological nature, analogous to the changes we present here for SDSS\,J1043+0855. Table\,\ref{t-ewtrip} lists the equivalent widths (subject to systematic uncertainties related to the method used in continuum fitting, as well as the statistical uncertainties given in Table\,\ref{t-ewtrip}) of the Ca\,{\textsc{ii}} emission lines in SDSS\,J1043+0855, which do not show any long term decay of the equivalent width of the Ca\,{\textsc{ii}} triplet such as seen at SDSS\,J1617+1620. Only SDSS\,J0738+1835 has displayed no changes in the shape and strength of the Ca\,{\textsc{ii}} triplet over a period of six years, although only three epochs are available, with two of them spaced only a year apart. 

\cite{manseretal16-1} showed that the variable Ca\,{\textsc{ii}} triplet line profiles of SDSS\,J1228+1040 could be interpreted as the emission from a fixed intensity pattern that precesses over a time scale of decades, possibly indicating a young debris disc that still has eccentric orbits and has not fully circularised. General relativistic precession will cause the debris to precess with a radially dependent period, causing orbits to cross one another and inducing collisions which produces the observed gaseous component to the debris disc. 

While the evolution of the emission from SDSS\,J1043+0855 appears to be remarkably similar to SDSS\,J1228+1040 and SDSS\,J0845+2257, the data have a lower signal to noise and have fewer epochs, and thus, while it is likely that the same physical mechanism is responsible for the evolution of the line profiles observed in all three systems, regular spectroscopic monitoring of all gas discs is necessary to develop a more detailed understanding of the dynamical processes present in planetary debris discs around white dwarfs.

Variability of debris discs is not only limited to the Ca\,{\textsc{ii}} triplet line profile. The dusty disc around SDSS\,J0959--0200 was observed to significantly decrease in infrared flux by \cite{xu+jura14-1}, who propose two mechanisms by which the disc could be depleted; a recent planetesimal impact on the disc, or instability near the inner edge. We suggest an additional scenario of a vertically extended cloud of dust, generated from an asteroid colliding with a pre-existing disc \citep{jura08-1}. Such an optically thin cloud would temporarily add to the infrared emission of the optically thick disc, but the overall infrared emission from the system would decrease as the dust cloud settled into the disc.

In Table\,\ref{t-wds} we also include four additional systems where circumstellar absorption of gaseous material has been detected around the host white dwarf, including WD\,1145+017, which is orbited by highly-dynamic debris, transiting the white dwarf with periods of $\simeq$\,4.5\,hr \citep{vanderburgetal15-1, xuetal16-1, gaensickeetal16-1, rappaportetal16-1}. WD\,1145+017 is unequivocally a highly dynamical and evolving system with a planetesimal currently undergoing disruption, and also hosts circumstellar gas absorption \citep{xuetal16-1}. Curiously, the detection of absorption due to circumstellar gas does not correlate with the presence of Ca\,{\textsc{ii}} triplet emission: SDSS\,1228+1040 is so far the only system in which both have been detected \citep{gaensickeetal12-1}. We note that circumstellar gas has been detected also around a number of hot and young white dwarfs, their origin is probably diverse in nature and not unambiguously associated with evolved planetary systems \citep{dickinsonetal12-2, barstowetal14-1}.

\section{Conclusions}
\label{sec:conc}

We report here the morphological variability of the Ca\,{\textsc{ii}} triplet in SDSS\,J1043+0855 on a time scale of nine years. The evolution of the Ca\,{\textsc{ii}} triplet reported here is similar to that of two other systems, SDSS\,J1228+1040 and SDSS\,J0845+2257. 

We have also analysed the optical spectra of SDSS\,J1043+0855 to determine its stellar parameters and the photospheric metal abundance. The Mg/Si and (upper limit to the) Fe/Si ratios of the planetary debris that has been accreted onto the white dwarf are broadly consistent with those of the crust of the Earth.

The recent detection of the 'real time' disruption of a planetesimal at WD\,1145+017, along with the dynamical evolution seen at the gaseous discs SDSS\,J1043+0855, SDSS\,J1228+1040, SDSS\,J1617+1620, and SDSS\,J0845+2257 reveals that variability at planetary systems around white dwarfs is more common than was initially thought. Additional spectroscopic and photometric monitoring of all the gaseous discs known so far is key to developing a more detailed understanding of the dynamical processes present in planetary debris discs at white dwarfs.

\section*{Acknowledgements}
The research leading to these results has received funding from the European Research Council under the European Union's Seventh Framework Programme (FP/2007-2013) / ERC Grant Agreement n.\,320964 (WDTracer). We would like to thank Yan-Ping Chen and Scott Trager for sharing their X-Shooter telluric template library. We thank the anonymous referee for a timely and constructive report.
 
Based on observations made with ESO Telescopes at the La Silla Paranal Observatory under programme IDs: 087.D-0139 and 386.C-0218. This work has made use of observations from the SDSS-III, funding for which has been provided by the Alfred P. Sloan Foundation, the Participating Institutions, the National Science Foundation, and the U.S. Department of Energy Office of Science. The SDSS-III web site is http://www.sdss3.org/. Based on observations made with the WHT operated on the island of La Palma by the Isaac Newton Group in the Spanish Observatorio del Roque de los Muchachos of the Instituto de Astrofísica de Canarias.

\bibliographystyle{mnras}
\bibliography{aamnem99,aabib}

\appendix
\newpage
\onecolumn

\section{Velocities and velocity gradients in an eccentric disc}\label{appendix-1}

\cite{horne+marsh86-1} show that line photons travelling through an optically thick disc are more likely to escape along paths of the greatest velocity gradient. We derive and list here the equations needed to calculate the radial velocity and velocity gradient at any point in an elliptical disc of constant orbital eccentricity, $e$ and constant orientation of its semi-major axes, $a$. This expands the formulation for modelling emission line profiles from circular discs described by \cite{horne+marsh86-1}. We first set up the basic components of an eccentric orbit, of which some are shown in Figure\,\ref{f-afig1}.

\subsection{Equations of an elliptical orbit}

The polar equation for an ellipse is given as

\begin{equation}\label{eq:polar}
r = \frac{l}{1 + e \cos \nu},
\end{equation}

\noindent where $l$ is the semi-latus rectum, and $\nu$ is the angle measured from the point of periastron, usually called the ``true anomaly''. Conservation of angular momentum in a central force field means that

\begin{equation}
r^2 \frac{\d \nu}{\d t} = h,
\end{equation}

\noindent where $h$, the specific angular momentum, is constant. This therefore leads to the true anomaly as a function of time being determined by the following equation

\begin{equation}
\frac{h t}{l^2} = \int_0^\nu \frac{\d \nu'}{\left(1 + e
  \cos \nu'\right)^2} .
\end{equation}

\begin{figure}
\centerline{\includegraphics[width=1\columnwidth]{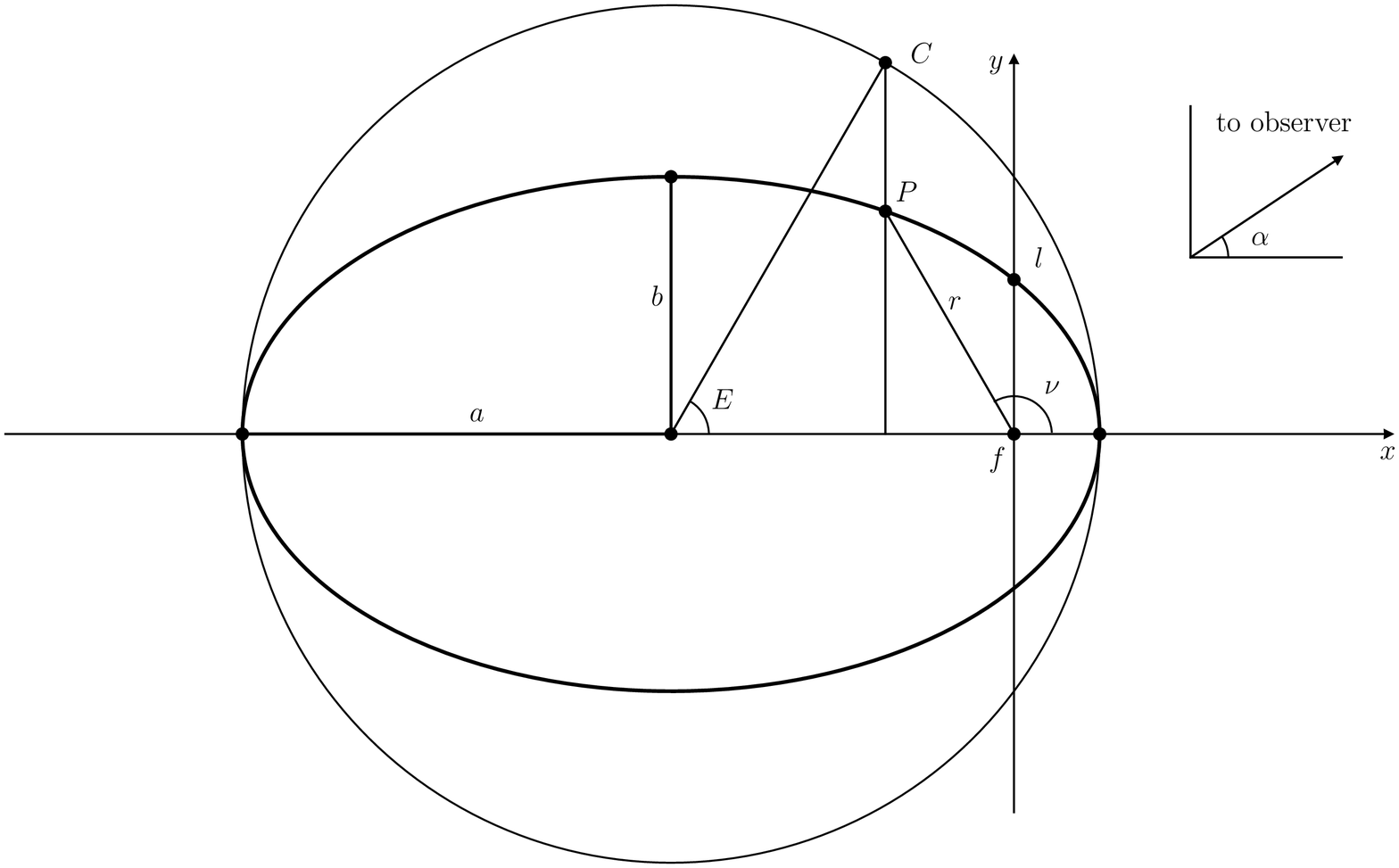}}
\caption{\label{f-afig1} An elliptical orbit of eccentricity, $e$, with the origin centred on the focal point, $f$, with semi-major and semi-minor axes $a$ and $b$. A point, $P$, along the orbit is shown, at a radius, $r$, and with a true anomaly, $\nu$. The semi-latus rectum, $l$, is also labelled, as is the eccentric anomaly, $E$, which can be constructed geometrically by putting a circle of radius, $a$, centred on the ellipse, and drawing a line perpendicular to the major axis through any point ($P$, here) on the ellipse. The angle, $\alpha$, to an observer is also shown, taken to be zero when the observer is looking along the $x$-axis from the positive side, and increases in the anti-clockwise direction.}
\end{figure}

\noindent Making the substitution $z = \tan \nu/2$ and setting

\begin{equation}
z = \left(\frac{1+e}{1-e}\right)^{1/2} \tan \frac{E}{2},
\end{equation}

\noindent where $E$ is the eccentric anomaly (see Figure\,\ref{f-afig1}), we get

\begin{equation}
E -  e \sin E  = \frac{h \left(1-e^2\right)^{3/2}}{l^2} t .
\end{equation}

\noindent When $\nu = 2\pi$, we get $E = 2\pi$ and $t = P$, so then we can write

\begin{equation}
E -  e \sin E  = n t, \label{eq:kepler}
\end{equation}

\noindent where $n$ is the mean angular velocity and is given by

\begin{equation}
n  = \frac{2\pi}{P} = \frac{h \left(1 - e^2\right)^{3/2}}{l^2}.
\end{equation}

\noindent Equation\,\ref{eq:kepler} is Kepler's equation of elliptical motion and can easily be solved numerically, and the true anomaly, $\nu$, follows from

\begin{equation}
\tan \frac{\nu}{2} = \left(\frac{1+e}{1-e}\right)^{1/2} \tan \frac{E}{2}.
\end{equation}

\noindent $E$ can be constructed geometrically by putting a circle radius $a$ centred on the ellipse, and drawing a line perpendicular to the major axis through any point on the ellipse. The angle subtended by the point on the circle where the perpendicular line cuts it is $E$. Using this, or the above equation, one can show that

\begin{equation}
\cos E = \frac{e + \cos \nu}{1 + e \cos \nu} .
\end{equation}

\noindent Since $E$ and $\nu$ always lie within the same range; $0$ to $\pi$ or $\pi$ to $2\pi$, this equation is sufficient to determine $E$. Related useful expressions are

\begin{eqnarray}
\cos \nu &=& \frac{\cos E - e}{1 - e \cos E}, \\
\sin \nu &=& \frac{\sqrt{1-e^2} \sin E}{1 - e \cos E}.
\end{eqnarray}

Assuming that the ellipse is oriented with the focus at the origin and the semi-major axis parallel to the $x$-axis (pointing left so that the periastron lies on the positive $x$-axis giving the standard orientation for $\nu$ in 2D polar coordinates, see Figure\,\ref{f-afig1}), then the velocity at any point is given by

\begin{eqnarray}
v_x & = & v_r \cos \nu - v_\nu \sin \nu, \label{eq:vx}\\
v_y & = & v_r \sin \nu + v_\nu \cos \nu,\label{eq:vy}
\end{eqnarray}

\noindent where 

\begin{equation}
v_r  = \frac{e l \sin \nu}{\left(1 + e \cos \nu\right)^2} \,
\dot{\nu},
\end{equation}

\noindent and

\begin{equation}
v_\nu  = \frac{l}{1 + e \cos \nu} \, \dot{\nu}, 
\end{equation}

\noindent where

\begin{equation}
\dot{\nu} = \left(\frac{1+e}{1-e}\right)^{1/2} \left(\cos^2 \frac{E}{2}
  + \frac{1+e}{1-e} \sin^2 \frac{E}{2}\right)^{-1} \, \dot{E},
\end{equation}

\noindent and 

\begin{equation}
\dot{E} = \frac{n}{1 - e \cos E} .
\end{equation}

\noindent The equation for $\dot\nu$ can be reduced to

\begin{equation}
\dot{\nu} = \frac{ \left(1-e^2\right)^{1/2}}{(1-e\cos E)^2} n,
\end{equation}

\noindent or, equivalently,

\begin{equation}
\dot{\nu} = \frac{ (1+e\cos \nu)^2}{\left(1-e^2\right)^{3/2}} n,
\end{equation}

\noindent giving the following alternative expressions for the two components of velocity (radial and azimuthal) 

\begin{equation}
v_r  = \frac{e n l \sin \nu}{(1 - e^2)^{3/2}}, \label{eq:vr}
\end{equation}

\noindent and

\begin{equation}
v_\nu  = \frac{n l (1 + e \cos \nu) }{(1 - e^2)^{3/2}}. \label{eq:vtheta}
\end{equation}

\noindent We are now in a position to consider orbits and radial velocities, and their derivatives.

\subsection{Velocity gradients in a confocal elliptical flow}

If a point in the disc is viewed by an observer at angle $\alpha$ defined so that $\alpha = 0$ when looking along the positive $x$-axis and increasing anti-clockwise (see Figure\,\ref{appendix-1}), then its radial velocity is given by

\begin{equation}
v_r = v_x \cos \alpha + v_y \sin \alpha .
\end{equation}

\noindent \cite{horne+marsh86-1} showed that the velocity gradient is also required to fully model the emission profile of an accretion disc, which using the notation $v_{,x} \equiv \frac{\partial v}{\partial x}$, is given by

\begin{eqnarray}
\frac{\d v_r}{\d k} &=& v_{r,x} \cos \alpha + v_{r,y} \sin \alpha , \\
 &=& 
v_{x,x} \cos^2 \alpha + \left(v_{x,y} + v_{y,x}\right)
\cos \alpha \sin \alpha +
v_{y,y} \sin^2 \alpha ,
\end{eqnarray}

\noindent where d$k$ is a line element towards the observer. From Equations\,\ref{eq:vx} \& \ref{eq:vy}, we have

\begin{eqnarray}
v_{x,x} &=& v_{r,x}\cos{\nu} - v_{\nu,x}\sin{\nu} - (v_r\sin{\nu} + v_\nu\cos{\nu})\nu_{,x},\\
v_{x,y} &=& v_{r,y}\cos{\nu} - v_{\nu,y}\sin{\nu} - (v_r\sin{\nu} + v_\nu\cos{\nu})\nu_{,y},\\
v_{y,x} &=& v_{r,x}\sin{\nu} + v_{\nu,x}\cos{\nu} + (v_r\cos{\nu} - v_\nu\sin{\nu})\nu_{,x},\\
v_{y,y} &=& v_{r,y}\sin{\nu} + v_{\nu,y}\cos{\nu} + (v_r\cos{\nu} - v_\nu\sin{\nu})\nu_{,y}.
\end{eqnarray}

\noindent One can show that

\begin{eqnarray}
\nu_{,x} &=& -\frac{y}{r^2},\\
\nu_{,y} &=& \frac{x}{r^2},
\end{eqnarray}

\noindent while, using Equations\,\ref{eq:vr} \& \ref{eq:vtheta} we obtain,

\begin{eqnarray}
v_{r,x}   &=& e(1-e^2)^{-3/2}[(n_{,x}l + nl_{,x})\sin{\nu} +nl\cos({\nu})\nu_{,x}],\\
v_{r,y}   &=& e(1-e^2)^{-3/2}[(n_{,y}l + nl_{,y})\sin{\nu} +nl\cos({\nu})\nu_{,y}],\\
v_{\nu,x} &=& (1-e^2)^{-3/2}[(n_{,x}l + nl_{,x})(1+e\cos{\nu}) - enl\sin({\nu})\nu_{,x}],\\
v_{\nu,y} &=& (1-e^2)^{-3/2}[(n_{,y}l + nl_{,y})(1+e\cos{\nu}) - enl\sin({\nu})\nu_{,y}].
\end{eqnarray}

\noindent From Equation\,\ref{eq:polar} we can obtain,

\begin{eqnarray}
l_{,x} &=& r_{,x}(1+e\cos{\nu}) - er\sin(\nu)\nu_{,x},\\
l_{,y} &=& r_{,y}(1+e\cos{\nu}) - er\sin(\nu)\nu_{,y},
\end{eqnarray}

\noindent where

\begin{eqnarray}
r_{,x} &=& \frac{x}{r},\\
r_{,y} &=& \frac{y}{r},
\end{eqnarray}

\noindent and from Kepler's third law, assuming the mass of the orbiting material to be negligible,

\begin{equation}
n^{2} = \frac{GM}{a^{3}},
\end{equation}

\noindent where $G$ is the gravitational constant and $M$ is the mass of the host star, we finally get the last unknowns,

\begin{eqnarray}
n_{,x} &=& -\frac{3l_{,x}n}{2l},\\
n_{,y} &=& -\frac{3l_{,y}n}{2l},
\end{eqnarray}

\noindent which now allows one to calculate the velocity derivatives.

\bsp

\label{lastpage}

\end{document}